\documentstyle[12pt]{article}
\def\beq{\begin{equation}}
\def\eeq{\end{equation}}
\newcommand{\bea}{\begin{eqnarray}}
\newcommand{\eea}{\end{eqnarray}}
\newcommand{\nn}{\nonumber}
\def\noi{\noindent}
\font\boldgreek=cmmib10
\mathchardef\mysigma="091B

\mathchardef\myalpha="090B

\begin{document}
\vbox to 3.5 truecm{}
\begin{flushright}
LPTHE-Orsay 95/15,\\
hep-ph/9504270
\end{flushright}
\begin{center}
{\large \bf MIXING AND CP VIOLATION IN D MESONS} \\
\vspace{6.0 ex}
{\large A. Le Yaouanc, L. Oliver, O. P\`ene and J.-C. Raynal} \\
\vspace{2.5ex}
{\large Presented by L. Oliver}\\
\vspace{2.5 ex}
{\large \it Laboratoire de Physique Th\'eorique et Hautes
Energies\footnote{Laboratoire associ\'e au
Centre National de la Recherche Scientifique - URA D0063}} \\
{\large \it Universit\'e de Paris XI, B\^atiment 211, 91405 Orsay Cedex,
France} \\
\vspace{3.5ex}
{\large Talk delivered at the Journ\'ees sur les projets \\
de Physique Hadronique, Soci\'et\'e Fran\c caise de Physique,\\
Super-Besse (France), 12-14 janvier 1995} \\
\end{center}
\vspace{4.5ex}
\noi {\bf R\'esum\'e}
\vspace{3.0ex}

Nous examinons le m\'elange et la violation de CP dans les m\'esons $D$
d'apr\`es le
Mod\`ele Standard, en soulignant les diff\'erences avec les autres m\'esons
pseudo-scalaires, et montrons
que les m\'esons $D$ peuvent \^etre utiles dans la recherche d'une nouvelle
physique au-del\`a du Mod\`ele
Standard.  \vskip 1 truecm
\noi {\bf Abstract}

\vspace{3.0ex}
We review mixing and CP violation in $D$ mesons, emphasizing the differences
with the other
pseudoscalar mesons in the Standard Model, and show that $D$ mesons can be
useful to look for physics
beyond the Standard Model.

\newpage
In the study of the weak interactions of pseudoscalar mesons, there are a
number of interesting
properties : i) decay rates~; ii) $P^0$-$\bar{P}^0$ mixing, and iii) CP
violation, that, if the
Standard Model is correct, result from the Cabibbo-Kobayashi-Maskawa matrix. It
is convenient to
parametrize this matrix following the phase convention and expansion in powers
of the Cabibbo angle
$\lambda = \sin \theta_C= 0.22$, due to Wolfenstein$^{1}$~:

\beq
V \cong \left ( \begin{array}{ccc}
1 - {\lambda^2 \over 2} &\lambda  &A\lambda^3(\rho - i\eta) \\
- \lambda  &1 - {\lambda^2 \over 2}  &A\lambda^2 \\
A\lambda^3 (1 - \rho - i\eta) &- A\lambda^2 &1 \\
\end{array} \right ) \ \ \ .
\eeq

The flavor structure of the Standard Model provides several interesting pairs
of neutral
pseudoscalar mesons $P^0$-$\bar{P}^0$, as indicated in Table 1. We indicate in
the Table the
power coun\-ting in terms of $\lambda$ of the dominant decay rates, mixing,
life-time differences and
CP asymmetries.  It is remarkable that this complex set of properties is for
the moment in
quantitative agreement with the expectations of the Standard Model, as
expressed by the matrix (1).
\par

	Mixing occurs through radiative corrections in the Standard Electroweak Theory
(box diagrams). The
mass eigenstates are :

\beq
|P_{1,2} > = p|P^0> \pm q|\bar{P}^0> \ \ \ .
\eeq

\noi This mixing produces $P^0 \leftrightarrow \bar{P}^0$ oscillations of
amplitude $e^{i\Delta Mt}$.
The parameter controlling the oscillations is $\Delta M/\Gamma$, given for the
different systems in
Table 1, where we also give $\Delta \Gamma/\Gamma$. \par

	Concerning CP violation, the unitarity of the Cabibbo-Kobayashi-Maskawa (CKM)
matrix implies, for
the different systems, triangular relations among CKM matrix elements. These
relations define
triangles in the complex plane that are of equal surface $S = A \lambda^6 \eta$
(due to the fact that
there is a single complex phase in the CKM matrix) but of different shapes
according to the considered
system (Table 1). The CP asymmetries are roughly proportional to the ratio
Surface of unitarity
triangle/Rate of the considered mode. It is remarkable that the simple power
counting gives the right
order of magnitude for the mixing parameters (within the present experimental
limits for
$D^0$-$\bar{D}^0$ and $B_s^0$-$\bar{B}_s^0$) and for the kaon CP
violation parameter $|\varepsilon | \sim 10^{-3}$.
\begin{center}
Table 1 \\
\vskip 3 mm
\begin{tabular}{|c|c|c|c|c|} \hline
& & & & \\
&$K^0$-$\bar{K}^0(\bar{s}d-s\bar{d})$	&$D^0$-$\bar{D}^0(\bar{u}c-c\bar{u})$
&$B_d^0$-$\bar{B}_d^0(\bar{b}d-b\bar{d})$
&$B_s^0$-$\bar{B}_s^0(\bar{b}s-b\bar{s})$ \\
& & & & \\\hline
Rates &$\lambda^2$ 	&1	&$\lambda^4$  &$\lambda^4$ \\ \hline
& & & & \\
${\Delta M \over \Gamma}$	&$m_c^2$  &$\lambda^2 m_s^2$ &$\lambda^2 m_t^2$
&$m_t^2$ \\
&$\tau \Delta M \cong 0.5$ &$\tau \Delta M < 0.08$ &$\tau \Delta M \cong 0.7$
&$\tau \Delta M > 9$ \\
\hline
& & & & \\
${\Delta \Gamma \over \Gamma}$ &1 &$\lambda^2$ &$\lambda^2$ &1  \\
&$\Gamma_S \gg \Gamma_L$ &$\tau \Delta \Gamma < 0.17$ & & \\
\hline
Unitarity &$\displaystyle{\sum_U} V_{Ud}V^*_{Us} = 0$ &$\displaystyle{\sum_D}
V_{Du} V^*_{Dc} = 0$
&$\displaystyle{\sum_{U}} V_{Ud} V^*_{Ub}= 0$ &$\displaystyle{\sum_U} V_{Us}
V^*_{Ub} = 0$ \\
triangles	& & & & \\ \hline
Surface of &$A^2\lambda^6 \eta$ &$A^2\lambda^6 \eta$	&$A^2\lambda^6 \eta$
&$A^2\lambda^6 \eta$ \\
triangles	& & & & \\ \hline
CP Asym. $\sim$ &$\lambda^4 \eta$ &$\lambda^6 \eta$	&$\lambda^2 \eta$
&$\lambda^2 \eta$ \\
${{\rm Surface} \Delta \over \Gamma {\rm (mode)}}$	&$|\varepsilon | \sim
10^{-3}$ & & & \\
& & & & \\ \hline
\end{tabular}
\end{center}

\par \vskip 5 mm
\section{\bf $D^0$-$\bar{D}^0$ mixing} \par

Charged $D^*$ decays can be used to identify the flavor of the neutral $D$ by
observing the pion
charge~: $D^{*^+} \to D^0 \pi^+$, $D^{*^-} \to \bar{D}^0 \pi^-$. Looking for
the time dependence in
the decay can help to separate$^{2}$ the process due to mixing $D^0 \to
\bar{D}^0 \to K^+ \pi^-$
from the doubly Cabibbo suppressed decay (DCSD) $D^0 \to K^+ \pi^-$. The time
dependence of the rate
is, for small time~:

\beq
R \left ( D^0 \to K^+ \pi^- \right ) \sim e^{-\Gamma t} \left [ |\rho_{DCS}|^2
+ {1 \over 2} (x^2
+ y^2) t^2 + \cdots \right ]
\eeq

\noi where $x = \Delta M/\Gamma$ , $y = \Delta \Gamma/2\Gamma$ and $\rho_{DCS}$
stands for the
amplitude of the doubly Cabibbo suppressed mode. The present limits obtained
with this method are
given in Table 1. \par

	As for the theory, the short distance box diagram (with the $b$ quark in the
loop) is highly
suppressed by powers of $\lambda$, and gives

\beq
\left ( {\Delta M \over \Gamma} \right )_{short \ distances} \sim 3 \times
10^{-5}	\ \ \ .
\eeq

\noi However, as pointed out by Wolfenstein$^{3}$, there are long distance
contributions, real
intermediate states of the type~:

\beq
D^0 \to  K\bar{K}, \pi \pi, \pi K, \pi \bar{K} \to \bar{D}^0 \ \ \ .
\eeq

\noi This sum vanishes in the exact SU(3) limit because of the GIM mechanism.
However, for some
individual modes, SU(3) is badly broken, like $\Gamma (K\bar{K}) \sim 3 \times
\Gamma (\pi \pi)$, and
Wolfenstein claims, as an order of magnitude, $(\Delta M/\Gamma)_{long \
distances} \sim
(\Delta \Gamma/\Gamma)_{long \ distances} \sim 0.01$. \par

	However, more detailed calculations of the dispersive part $\Delta M/\Gamma$
with the intermediate
states (4) by Donoghue et al.$^{4}$ and updated by Pakvasa$^{5}$, and also QCD
symmetry
arguments on the operators that can contribute to $D^0$-$\bar{D}^0$ mixing by
Georgi$^{6}$, point
rather to a much smaller result for the long distance contribution, at most of
the order :

\beq
\left ( {\Delta M \over \Gamma} \right )_{long \ distances} \sim 10 \times
\left ( {\Delta M
\over \Gamma} \right )_{short \ distances}	\ \ \ .	 \eeq

\noi If this estimation is the correct one, it leaves room for the search of
physics beyond the
Standard Model, as shown in Table 2$^{5}$. However, at least in the case of
$\Delta
\Gamma/\Gamma$, it remains to be seen how Georgi symmetry arguments are
realized concretely in the sum
over real intermediate states, where Wolfenstein's argument seems on firm
ground.

\begin{center}
Table 2 \\
\vskip 3 mm

\begin{tabular}{|l|c|} \hline
&$\Delta M/\Gamma$ \\ \hline
Mod\`ele Standard	&$10^{-5} - 10^{-4}$ \\ \hline
Two Higgs model	 &$>10^{-4} (\tan \beta \gg 1$) \\ \hline
4th generation	&$10^{-2}$ \\ \hline
Flavor-changing neutral Higgs &$10^{-2} -10^{-1}$ \\ \hline
SUSY	&$10^{-5} - 10^{-4}$ \\ \hline
L-R Symmetry	&$10^{-5} - 10^{-4}$ \\ \hline
\end{tabular}
\end{center}

\vskip 5 mm
\section{\bf CP violation}
\subsection{Standard Model}

As we see in Table 1, the CP asymmetries are very small in the Standard Model
for Cabibbo allowed $D$
decays (naively one expects numbers of the order or smaller than $10^{-5}$).
\par

	However, in Cabibbo suppressed decays like $D \to \pi \pi$, $K\bar{K}$,
$K\bar{K}^*$, $\rho \pi$, ...
one can have higher asymmetries from the interference between the tree
amplitude, the Penguin diagram
amplitude, and strong phases coming from nearby resonances. The CP asymmetries
can occur because the
tree amplitude $(\sim V_{ud} V_{cd}^*)$ has a different phase than the Penguin
diagram $(\sim~V_{ub}
V_{cb}^*)$ coming from the operator induced by radiative corrections with the
$b$ quark in the loop~:

\beq
- {\alpha_s(\mu^2) \over 12\pi} \log \left ( {m^2_b \over \mu^2} \right )
V_{ub}V_{cb}^* \left [ \bar{u} \gamma_{\mu}(1 - \gamma_5) \lambda^a  c \right ]
\left [ \bar{q} \gamma_{\mu} \lambda^a  q \right ]	 \ \ \ . \eeq

\noi For example$^{7}$, considering for the mode $D \to K\bar{K}^*$ there are
resonances
near the $D$ mass that provide the necessary strong phases.

\beq
A \left ( D \to K\bar{K}^* \right ) = V_{ud}V_{cd}^* \ A_{tree} +
V_{ub}V_{cb}^* \ A_{Penguin} \ \ \ .		 \eeq

\noi There are two isospin amplitudes $\Delta I=1/2$ (tree and Penguin) and
$\Delta I=3/2$ (tree),
resulting in an asymmetry that can occur even for charged $D$ mesons (direct CP
violation) :

\beq
Asym \left ( D \to K \bar{K}^* \right ) \sim Im \left ( V_{ud} V_{cd}^* V_{cb}
V_{ub}^* \right ) Im
\left [ A_{3/2} (A_{1/2})^* \right ]	 \ \ \ . \eeq

\noi One finds rather large asymmetries, of the order $Asym (K\bar{K}^*) \sim
10^{-3}$ while for other
modes, the asymmetry is smaller~: $Asym(K^+K^-) \sim 10^{-4}$ (the present
experimental
limit$^{8}$ is $A(K^+K^-) < 0.45$).

\vskip 5 mm
\subsection{Beyond the Standard Model}

Since CP asymmetries are expected to be very small in Cabibbo allowed decays,
these could be the
right place to look for other possible sources of CP violation beyond the
Standard Model$^{9}$.
\par

	As an example, let us consider the decay mode $D \to \bar{K}\pi$ where we have
two isospin channels,
$\Delta I=1/2$ and $\Delta I=3/2$. The Final State Interaction phases are
expected to be large because there is a nearby wide resonance $K_0^*(1950)$
coupled to these channels.
On the other hand, as emphasized by Bauer, Stech and Wirbel$^{10}$, one needs a
large phase shift
$\delta_{1/2} - \delta_{3/2} = (77 \pm 11)^{\circ}$ to account for the ratio
$\Gamma (D^0 \to \bar{K}^0 \pi^0)/\Gamma (D^0 \to K^- \pi^+)$ that otherwise
would be too small. \par

There are no Penguins in these decays, and only the tree amplitude $\sim V_{ud}
V_{cs}^*$
contributes in the Standard Model. Therefore, from the weak interaction point
of view, the amplitudes
$\Delta I=1/2$ and $\Delta I = 3/2$ have the same phase, leading to a vanishing
direct CP asymmetry.
However, there could be a very small asymmetry from mixing by interference
between the amplitudes
$D^0 \to K^- \pi^+$ and $D^0 \to \bar{D}^0 \to  K^- \pi^+$, leading to an
asymmetry of the
order $\sim 10^{-6}$. \par

	If there is another source of CP violation beyond the Standard Model, in
general the weak phases of
the amplitudes $\Delta I = 1/2$ and $\Delta I = 3/2$ will be different, and we
will have an asymmetry

\bea
Asym (K \pi) &\sim &\sin \left ( \delta_{1/2} - \delta_{3/2} \right )
\left [ \left ( M_{1/2}^{SM} \right )^* M_{3/2}^{BSM} -
\left ( M_{3/2}^{SM} \right )^* M_{1/2}^{BSM} \right ] \nn \\ &\sim &4 \times
10^{-2}
\sin \varphi_{BSM}  {(TeV)^2 \over \Lambda^2}				 \eea

\noi where $\varphi_{BSM}$ is a CP phase coming from physics beyond the
Standard Model. The asymmetry
will test this new source of CP violation if $\sin \varphi_{BSM} \gg
\lambda^4$. \par

	To observe an asymmetry of $O(10^{-2})$ at the $3\sigma$ level one needs $10^6
\ D^0\bar{D}^0$ pairs,
a number that can be reached at a Tau-Charm Factory. \par

	However, ``realistic'' models of CP violation beyond the Standard Model,
constrained by the kaon CP
parameters $\varepsilon$ and $\varepsilon '$, predict a much weaker asymmetry.
For example, in the
L-R symmetric model with spontaneous CP violation one expects an asymmetry
$\leq 2 \times 10^{-4}$.
Still, the charm sector could be enchanced by some unknown reason (Higgs
couplings~?) and it is
worth to look for CP violation in these modes.

section*{Acknowledgements}

We would like to thank the Clermont-Ferrand team who has so much contributed to
the success of this meeting.
This work was supported in part by the CEC Science Project SC1-CT91-0729 and by
the Human Capital and Mobility Programme, contract CHRX-CT93-0132.

\noi 

\begin{thebibliography}{11}

\bibitem{} L. Wolfenstein, Phys. Rev. Lett. {\bf 51}, 1945 (1983).
\bibitem{} J. C. Anjos, Phys. Rev. Lett. {\bf 60}, 1239 (1988).
\bibitem{} L. Wolfenstein, Phys. Lett. {\bf B164}, 170 (1985).
\bibitem{} J. F. Donoghue et al., Phys. Rev. {\bf D33}, 179 (1986).
\bibitem{} S. Pakvasa, Charm as probe of new physics, Invited talk at CHARM
2000 Workshop, Batavia,
Ill. (1994), Univ. of Hawaii preprint UH-511-787-94.
\bibitem{} H. Georgi, Phys. Lett. {\bf B297}, 353 (1992).
\bibitem{} F. Buccella et al., Phys. Lett. {\bf B302}, 319 (1993).
\bibitem{} J. C. Anjos {\it et al.}, Phys. Rev. {\bf D44}, R3371 (1991).
\bibitem{} A. Le Yaouanc, L. Oliver and J.-C. Raynal, Phys. Lett. {\bf B292},
353 (1992).
\bibitem{} M. Wirbel, B. Stech and M. Bauer, Z. Phys. {\bf C29}, 637 (1985)~;
{\bf C34}, 103
(1987).   \end{thebibliography}
\end{document}